\documentclass[a4paper,11pt,twocolumn]{article}

\usepackage{icphs2023}
\usepackage{metalogo} 
\usepackage{epstopdf}
\usepackage{verbatim}

\usepackage{booktabs}
\usepackage{multirow}
\usepackage[hidelinks]{hyperref}
\usepackage{amsmath}
\usepackage{wasysym}

\title{Automatic Sound Event Detection and Classification of Great Ape Calls Using Neural Networks}
\author{Zifan Jiang$^{1,2}$, Adrian Soldati$^{1,3}$, Isaac Schamberg$^2$, Adriano R. Lameira$^4$, Steven Moran$^{1,5}$}
\organization{$^1$University of Neuch\^atel, $^2$University of Zurich, $^3$University of St Andrews,\\ \vspace{-.4\baselineskip}$^4$University of Warwick, $^5$University of Miami}
\email{jiang@cl.uzh.ch, \{adrian.soldati, steven.moran\}@unine.ch, isaac.schamberg@uzh.ch, adriano.lameira@warwick.ac.uk}
\begin{document}

\maketitle

\begin{abstract}
We present a novel approach to automatically detect and classify great ape calls from continuous raw audio recordings collected during field research. Our method leverages deep pretrained and sequential neural networks, including wav2vec 2.0 and LSTM, and is validated on three data sets from three different great ape lineages (orangutans, chimpanzees, and bonobos). The recordings were collected by different researchers and include different annotation schemes, which our pipeline preprocesses and trains in a uniform fashion. Our results for call detection and classification attain high accuracy. Our method is aimed to be generalizable to other animal species, and more generally, sound event detection tasks. To foster future research, we make our pipeline and methods publicly available.\footnote{\url{https://github.com/J22Melody/sed\_great\_ape}}
\end{abstract}




\section{Introduction}

In primatology, as in documentary linguistics, the collection, annotation, and analysis of primary field data are time-consuming and expensive. The recordings of primate calls are also often undertaken in not ideal environmental conditions (e.g., in a dense and noisy forest where other vocal species are present), making the process even more challenging. Therefore it is typical that primatologists first manually annotate their recordings and then conduct acoustic analyses on their species' specific calls. 

Our goal in this paper is to automatically and accurately detect and classify primate calls from raw audio data. After discussing related work (\S\ref{sec:related_work}), we describe our data sets (\S\ref{sec:data}) and present our method that leverages pretrained and sequential neural networks (\S\ref{sec:method}). 
We carry out reliable and reproducible experiments (\S\ref{sec:experiments}) to support the effectiveness of our approach and compare our results between different architectural choices on the data sets of three great ape lineages. Overall, our models achieve more than 80\% frame-level classification accuracy and weighted F1-score.
Interestingly, we find that the wav2vec 2.0 \cite{baevski2020wav2vec} model -- even though pretrained on a large corpus of human speech \cite{panayotov2015librispeech} -- generalizes surprisingly well as an audio feature representation layer for great ape calls without additional fine-tuning. This finding perhaps bides well with the similar vocal tract morphology between extant great apes and humans, and the larger prediction that ancestral ape-like calls evolved to become the building blocks of human speech. 


\begin{table*}[!ht]
\footnotesize
\centering
\begin{tabular}{lrrrr}
\toprule
\textbf{data set} & \textbf{\# audio clips} & \textbf{mean / call / total duration} & \textbf{\# indiv. (\male / \female)} & \textbf{\# call types (duration ratio) / units} \\
\midrule

chimpanzee & 235 & \textasciitilde\, 8s / 1,955s / 1,964s & 11 (11 / 0) & 4 (6:3:3:1) / 686 \\

orangutan & 65 & \textasciitilde\, 74s / 2,793s / 4,817s & 10 (10 / 0) & 7 (700:40:200:100:20:1:1) / 9016 \\

bonobo & 28 & \textasciitilde\, 24s / 62s / 677s & 7 (3 / 4) & 18 (20:40:10:20:...:200:...:5:1) / 356 \\

\bottomrule
\end{tabular}
\caption{Data set overview and stats. Shown in the table are the number of audio clips, mean duration of each clip, total duration of calls of all clips, total duration of all clips, number of individuals (male and female), number of call types (approximate total duration ratio of them, partially omitted for bonobo due to space limit), and number of individually annotated call units.}
\label{tab:data_stats}
\end{table*}

\section{Related Work}\label{sec:related_work}


To the best of our knowledge, there is no precedent research on the automatic detection and classification of great ape calls from raw audio. Therefore, we find our work lies in between general-purpose sound event detection and human speech recognition. We briefly discuss these also in light of existing research on animal sound classification.

\textbf{Sound Event Detection (SED)}
aims at automatically recognizing \textit{what} is happening in an audio signal and \textit{when} it is happening \cite{mesaros2021sound}. SED tasks are usually general-purpose (e.g., birds singing or footsteps) as opposed to domain-specific tasks like human speech or music analysis, and thus encounter particular challenges, such as the \textit{cocktail party effect} \cite{arons1992review}. Notably, the DCASE challenge that took place in recent years involves relevant SED tasks \cite{Turpault2019_DCASE}. 
A variant of SED is the audio tagging task \cite{Mesaros2018_TASLP}, in which only \textit{what} is happening in an audio signal is annotated and recognized, but not \textit{when}. 

Prominent work on both tasks \cite{ebbers2021self, gong2021ast} tends to consist of common components, including: 
(1) an acoustic feature representation layer, either 
(a) traditional spectrogram-based approaches or 
(b) convolutional neural networks (CNN) on spectrogram features or on raw waveforms \cite{dai2017very};
(2) a sequence modeling layer that captures temporal interaction between frame-level features, which takes the forms of mean/max pooling strategies, recurrent neural networks (RNN) or Transformers \cite{vaswani2017attention};
and (3) an objective function, usually cross-entropy classification on frame-level (for SED) or clip-level (for audio tagging), or occasionally CTC \cite{graves2006connectionist} for ``sequentially labeled data'' \cite{hou2018polyphonic}.


\textbf{Animal sound classification}
involves related research that addresses animal sounds \cite{fedurek2016sequential, sun2021classification, recalde2023pykanto}. While this line of research uses similar acoustic feature representation techniques as seen in SED tasks, the models tend to work on top of individual short audio units of animal vocalizations that are manually selected from raw audio recordings by human experts, and thus fall short on dealing with a continuous audio signal that contains many audio events as well as noise. Instead, our method detects and classifies great ape calls directly from raw recordings ranging from seconds to minutes (theoretically recurrent models also extend to hours, but it could pose challenges in model training time).

\textbf{Automatic Speech Recognition (ASR)} on humans -- one of the extant five great apes -- has seen significant progress \cite{li2022recent} (while animal calls remain mysterious to fully decipher), including the recent work wav2vec 2.0 \cite{baevski2020wav2vec}. 
We are interested in whether the wav2vec 2.0 speech representation pretrained on Librispeech \cite{panayotov2015librispeech} can be applied successfully to great ape call detection and classification.

\section{Data}
\label{sec:data}


Our data consists of recordings of chimpanzee pant-hoots, orangutan long calls, and bonobo high-hoots. Table \ref{tab:data_stats} provides an overview of these data sets.


\textbf{Chimpanzee pant-hoots} are vocal sequences composed of up to four acoustically distinct phases, typically produced in this order: \textit{introduction}, \textit{build-up}, \textit{climax}, and \textit{let-down} \cite{marler1975individuality}. 
Most pant-hoots contain two or more phases, although single phases can be produced in specific contexts \cite{soldati2022audience}. The successive nature and well-balanced phase duration facilitate the training of a classifier (\S\ref{sec:exp_chimp}). We use annotated recordings of isolated pant-hoots (i.e., no temporal overlap with others' calls) produced during feeding, traveling, and resting context.





\textbf{Orangutan long calls} are composed by a \textit{full pulse}, which is sub-divided into a \textit{sub-pulse transitory element} and a \textit{pulse body}, or into sequences of \textit{bubble sub-pulse} or \textit{grumble sub-pulse} \cite{hardus2009description, lameira2008orangutan}. The complex temporal interrelationship between phases and the overlapping and class-unbalanced characteristics (some phases are extremely short) pose challenges for training a multi-class classifier. On the other hand, the duration of calls and non-calls in the data set are well balanced, which opens the gate to a binary call detection model (\S\ref{sec:exp_orang}). Annotated recordings include spontaneous long calls and long calls produced in response to other males' vocal presence or environmental disturbances. 





\textbf{Bonobo high-hoots} -- which make up the plurality of our bonobo call data set -- are loud, tonal vocalizations \cite{schamberg2016call}. Unlike the chimpanzee and orangutan data sets, call types other than the homologous loud calls are also present in the data set, but are rather minor and diverse (e.g., peep-yelp, soft barks), which could be challenging to find. The total call duration is relatively short.





\section{Method}
\label{sec:method}

This section introduces our method, which is illustrated in Fig.~\ref{fig:method}. We describe each step in turn.

\subsection{Audio Preprocessing}
First, all audio clips are converted to \textit{.wav} format and resampled to 16 kHz sample rate. We segment them into 20ms frames and pad with zeros if necessary.


\subsection{Feature and Label Extraction}
Next, we extract three types of acoustic features with the \textit{torchaudio} Python package. For an audio clip of $T$ frames, we get a sequence of frame-level features $I_{1:T}$, in the shape of $T \times feature\_dim$:


\begin{itemize}
\item \textbf{Raw waveform}: the original amplitude values over time. $feature\_dim = 0.02 \times 16000 = 320$.

\item \textbf{Spectrogram}: calculated then from the raw waveform with $feature\_dim = 201$.

\item \textbf{wav2vec 2.0}: inferred from the raw waveform by the \textit{WAV2VEC2\_BASE} model on a CPU device with $feature\_dim = 768$.
\end{itemize}

\noindent We then extract frame-level labels from our annotations. For each annotated unit, we mark all frames inside the annotated time span with a positive class index indicating the call type (e.g., 1 for \textit{intro}); unmarked frames are 0s by default, indicating non-calls. The labels' shape of a clip $L_{1:T}$ is $T \times 1$.

\subsection{Data Split}

We shuffle the clips (features and labels) of each data set and split them into 80\%, 10\%, and 10\% for training, validation, and test sets, respectively. We make three different splits by three different random seeds 0, 42, and 3407 \cite{picard2021torch} to allow multiple experiments and to study the effect of randomness.

\subsection{Sequence Modeling}
\label{sec:sequence}

Since our goal is to learn a function $f$ from $I_{1:T}$ to $L_{1:T}$, we first input $I_{1:T}$ to a sequence modeling function $f^m$ that outputs a hidden sequence $H_{1:T}$ ($T \times hidden\_dim$), which captures inter-frame interaction. $H_{1:T}$ then goes through a dense linear function $f^d$ (output $D_{1:T}$) followed by a Softmax function $f^s$ to produce the probability distribution over target classes, i.e., $P_{1:T}$ ($T \times num\_class$). 

Specifically, $f^m$ takes the following forms:

\begin{itemize}

\item \textbf{RNN}: a (bidirectional) LSTM with hidden size 1024. $hidden\_dim = 1024 * 2 = 2048$.

\item \textbf{Transformer encoder}: a Transformer encoder with 8 heads and 6 layers. $hidden\_dim = 1024$.

\item \textbf{Autoregressive model}: we optionally add autoregressive connections between time steps to encourage consistent output labels. The output of $f^d$ at time step $t$, i.e., $D_t$ is concatenated to the next time step's input $I_{t+1}$. By default, it indicates that the RNN layer goes unidirectionally, but one can also stack two autoregressive layers and add them up before the Softmax operation to retain bidirectionality.

\end{itemize}

Lastly, a cross-entropy loss is computed between the model output $P_{1:T}$ and the gold labels $L_{1:T}$, then backpropagated to train the models. To mitigate the class imbalance problem, we set class weights to the reciprocal of each class's occurrence times. 

\begin{figure}[!t]
\begin{center}
\includegraphics[width=\linewidth]{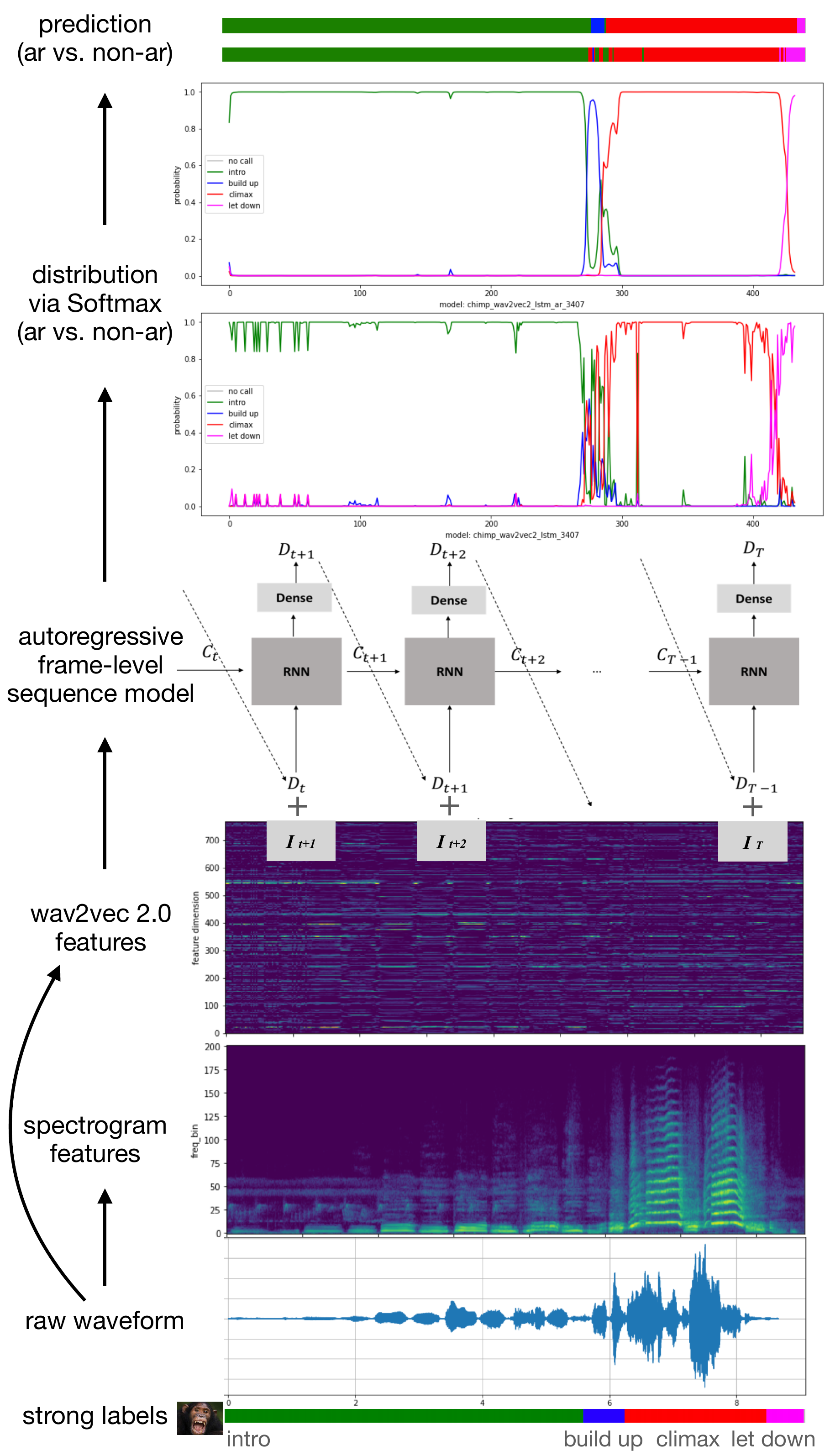}
\caption{Our method (bottom-up). The audio clip shown in the figure is from the chimpanzee data set. \textit{(Non-)ar} stands for (non-)autoregressive.}\label{fig:method}
\end{center}
\end{figure}

\begin{table*}[!ht]
\footnotesize
\centering
\begin{tabular}{llllrrrrr}
\toprule
\textbf{ID} & \textbf{data} & \textbf{feature} & \textbf{model} & \textbf{dev acc.}& \textbf{dev f1} & \textbf{test acc.} & \textbf{test f1}	& \textbf{aucpr} \\

\midrule
\multicolumn{8}{l}{\textit{Explore the best feature and model combination}}\\

\rule{0pt}{2ex}E1 & chimp & waveform & lstm (baseline) & $51.7\pm2.1$ & $35.4\pm2.5$ & $51.0\pm3.6$ & $34.7\pm4.3$ & - \\ 
E1.1 & chimp & spectrogram & lstm & $60.3\pm1.5$ & $55.7\pm2.3$ & $58.7\pm4.7$ & $53.9\pm5.4$ & - \\ 
E2 & chimp & wav2vec2 & lstm & $81.0\pm4.0$ & $79.9\pm4.6$ & $79.3\pm2.3$ & $77.9\pm3.6$ & - \\ 
E2.1 & chimp & wav2vec2 & transformer & $71.3\pm0.6$ & $68.3\pm0.2$ & $75.3\pm0.6$ & $72.1\pm0.5$ & - \\ 

\midrule
\multicolumn{8}{l}{\textit{Explore the hyper-parameters}}\\

\rule{0pt}{2ex}E3.1 & chimp & wav2vec2 & lstm (E2 + $batch\_size=4$) & $69.7\pm1.5$ & $71.8\pm2.6$ & $67.7\pm4.0$ & $69.6\pm4.0$ & - \\ 
E3.2 & chimp & wav2vec2 & lstm (E2 + $batch\_size=8$) & $63.3\pm0.6$ & $62.6\pm1.0$ & $62.0\pm4.4$ & $61.5\pm4.0$ & - \\ 
E3.3 & chimp & wav2vec2 & lstm (E2 + $dropout=0.2$) & $80.7\pm3.5$ & $80.0\pm4.4$ & $78.0\pm1.7$ & $76.8\pm2.7$ & - \\ 
E3.4 & chimp & wav2vec2 & lstm (E2 + $dropout=0.1$) & $81.0\pm4.0$ & $80.2\pm4.8$ & $78.7\pm2.9$ & $77.3\pm3.9$ & - \\ 
E3.5 & chimp & wav2vec2 & lstm (E2 - $balance\_weights$) & $81.0\pm3.6$ & $79.6\pm4.4$ & $79.3\pm2.3$ & $78.3\pm3.6$ & - \\ 

\midrule
\multicolumn{8}{l}{\textit{Explore autoregressive modeling}}\\

\rule{0pt}{2ex}E4 & chimp & wav2vec2 & lstm (E2 + autoregressive) & $87.7\pm1.2$ & $87.1\pm1.8$ & $85.7\pm2.1$ & $85.6\pm2.5$ & - \\ 

\midrule
\multicolumn{8}{l}{\textit{Extend to orangutan long calls and a binary setting}}\\

\rule{0pt}{2ex}E5 & orang & wav2vec2 & lstm (= E4) & $83.0\pm1.0$ & $82.7\pm1.4$ & $81.7\pm3.1$ & $82.0\pm2.6$ & - \\ 
E5.1 & orang & wav2vec2 & lstm (E5 + binary target) & $92.3\pm2.5$ & $92.1\pm2.5$ & $92.0\pm1.0$ & $91.9\pm1.1$ & $0.96$ \\ 

\midrule
\multicolumn{8}{l}{\textit{Extend to bonobo calls and a binary setting}}\\

\rule{0pt}{2ex}E6 & bonobo & wav2vec2 & lstm (= E4) & $87.0\pm4.6$ & $85.9\pm6.3$ & $83.7\pm3.8$ & $82.3\pm2.2$ & - \\ 
E6.1 & bonobo & wav2vec2 & lstm (E6 + binary target) & $92.0\pm3.6$ & $91.9\pm3.4$ & $87.7\pm3.5$ & $87.8\pm2.9$ & $0.87$ \\ 

\midrule
\multicolumn{8}{l}{\textit{Zero-shot transferring from orangutan to bonobo}}\\

\rule{0pt}{2ex}E7 & bonobo & wav2vec2 & lstm (= E5.1) & $63.0\pm13$ & $69.2\pm10$ & $72.0\pm4.0$ & $74.2\pm3.1$ & $0.55$ \\ 

\bottomrule
\end{tabular}
\caption{Experimental results. We run all experiments three times based on different random seeds and report the mean and standard deviation. \textit{acc.} stands for frame-level accuracy, \textit{f1} stands for the frame-level average F1-score weighted by the number of true instances per class, and \textit{aucpr} stands for the area under the precision-recall curve for the positive class in the binary case at test time when the random seed is set to 0. For hyper-parameters, we start \textit{E1} with $batch\_size=1$, $dropout=0.4$ and keep them by default, if not otherwise specified in the table.}
\label{tab:results}
\end{table*}

\section{Experiments and Results}\label{sec:experiments}

Our experiments were done in PyTorch \cite{paszke2019pytorch}, with Python 3.8 on an Nvidia Tesla V100 GPU (32GB ram). Most models have $\sim$40k million parameters and finish the training of up to 200 epochs with early stopping on validation F1-score within one hour (or a few hours because autoregressive recurrent models run slowly on PyTorch). Table~\ref{tab:results} presents our results.

\subsection{Initial Exploration with Chimpanzee Data}
\label{sec:exp_chimp}


We first test the viability of our approach on the chimpanzee data in light of the simplicity of the pant-hoot annotation scheme. 
We start from \textit{E1}, a simple waveform + LSTM baseline, and observe in \textit{E2} that wav2vec 2.0 outperforms the raw waveform and spectrogram by a large margin, which demonstrates the power of transfer learning from pretraining on human speech. 
In \textit{E2.1}, we find that the Transformer encoder does not outperform LSTM. Hence, we infer that Transformer's ability to capture arbitrary long-range dependency is not beneficial to our task and/or our small data sets. 
Next, we explore the hyper-parameters in the second experimental group and we find in \textit{E3.5} that balancing class weights (\S\ref{sec:sequence}) has a small impact. 
Lastly, we show in \textit{E4} that the autoregressive connections are beneficial for consistent output, as illustrated by the tiny gaps in \textit{non-ar} output in Fig.~\ref{fig:method}.

\subsection{Extending to Orangutan and Bonobo Data}
\label{sec:exp_orang}




We successfully extend the model in \textit{E4} to as trained on the orangutan (\textit{E5}) and bonobo (\textit{E6}) data sets. We note that some minority classes perform less well due to data scarcity, in contrast to the well-balanced situation in the chimpanzee data set (see Table~\ref{tab:data_stats}).

We further reduce the task to a binary (call vs.\ non-call) classification that resembles voice activity detection of human speech. This is a useful tool to automatically extract calls from raw recordings for further bioacoustic analysis (e.g., studying the repertoire of a given species).

Finally, to understand the generalizability of our models, we try zero-shot transferring the model trained in \textit{E5.1} on orangutans directly to unseen bonobo data. The results show that it is promising to build a potential general-purpose sound event detection model for all great ape calls.

\section{Discussion}
We have addressed a gap in the bioacoustics research of non-human great apes by developing an approach for automatically identifying sound events and classifying great ape calls using a neural network architecture. Our method successfully and accurately identifies and classifies calls in three species of non-human great apes, and provides a tool for primatologists to bootstrap call identification and analysis from raw unannotated audio recordings. 

Our method also shows the general applicability of the wav2vec 2.0 model trained on human speech for identifying vocalizations and call types in other species. Future work may apply our approach to more animals, as part of the goal to decode the communication systems of great apes and other non-human animals more broadly.\footnote{For example: \url{https://www.earthspecies.org}.}

\theendnotes

\section{Acknowledgements}
We gratefully acknowledge funding from: UK Research \& Innovation, Future Leaders Fellowship (MR/T04229X/1; ARL), Swiss National Science Foundation (PCEFP1\_186841: SM, ZJ; 310030\_185324: AS), St Leonard College (AS), and Swissuniversities (AS). We thank also Klaus Zuberb{\"u}hler, Josep Call, and the field assistants of the Budongo Conservation Field Station.


\bibliographystyle{IEEEtran}
\bibliography{icphs2023}

\end{document}